# Integration of SCADA services in cross-infrastructure holistic tests of cyber-physical energy systems

Van Hoa Nguyen
Univ. Grenoble Alpes, INES,
CEA, LITEN, Department of Solar Technologies,
F-73375 Le Bourget du Lac, France
vanhoa.nguyen@cea.fr

Tung Lam Nguyen
Univ. Grenoble Alpes, G2Elab
CNRS, G2Elab
F-38000 Grenoble, France
tung-lam.nguyen@grenoble-inp.fr

Quoc Tuan Tran
Univ. Grenoble Alpes, INES,
CEA, LITEN, Department of Solar Technologies
F-73375 Le Bourget du Lac, France
quoctuan.tran@cea.fr

Yvon Besanger
Univ. Grenoble Alpes, G2Elab
CNRS, G2Elab
F-38000 Grenoble, France
yvon.besanger@g2elab.grenoble-inp.fr

Raphael Caire
Univ. Grenoble Alpes, G2Elab
CNRS, G2Elab
F-38000 Grenoble, France
raphael.caire@g2elab.grenoble-inp.fr

*Abstract*— Cyber-Physical Energy System, due to its multi-domain nature, requires a holistic validation methodology, which may involve the integration of assets and expertise from various research infrastructures. In this paper, the integration of Supervisory Control and Data Acquisition services to cross-infrastructure experiment is proposed. The method requires a high degree of interoperability among the participating partners and can be applied to extend the capacity as well as the degree of realism of advanced validation method such as co-simulation, remote hardware-in-the-loop or hybrid simulation. The proposed method is applied to a case study of multi-agent system based control for islanded microgrid where real devices from one platform is integrated to real-time simulation and control platform in a distanced infrastructure, in a holistic experimental implementation.

## I. INTRODUCTION

In response to the requirements of providing better support for a higher penetration of Distributed Renewable Energy Sources (DRES) and introducing a wide range of beyond state of the art applications based on integration of Information and Communication Technologies (ICT), the traditional power system has been evolving itself into a multi-layered Cyber-Physical Energy System (CPES) (i.e. smart grid) – a juxtaposition of technologies with fundamentally different natures [1]. Due to the complex interactions and interdependency in CPES, it is mandatory to develop a comprehensive and holistic validation strategy that the classical component testing methods can no longer supply [2].

The development of such a comprehensive and holistic validation strategy for CPES faces several challenges: most importantly, the ICT systems and the energy systems (electrical grid, heating system) feature distinct behaviours (continuous - discrete event), and need to be considered in various different time scales, from several µs (electrical transient behaviour) to several days (market management). A holistic validation of CPES therefore requires a suitable complexity of validation environment which is not always feasible in the framework of a single Research Infrastructure (RI). The correctness and close to reality of models of DRES and ICT in simulation are also requirements that are difficult to satisfy due to the continuously and actively developing environment. Researchers have come up with some innovative approaches to overcome these challenges: e.g. co-simulation to investigate the interconnection of domains from a global point of view [3]–[5], hardware-in-the-loop (HIL) where a real hardware is integrated to the validation framework to consider the realistic behaviour of such hardware and its impact to the system [6], [7], hybrid-simulation [8], integration of HIL to co-simulation [9]. These approaches can be implemented either in a single platform or by coupling several complementary platforms in a cross-infrastructure manner [10], [11].

As aforementioned, a holistic validation framework for CPES requires a certain degree of complexity that sometimes exceeds the capacity of one single infrastructure in terms of hardware equipment and in terms of staff's expertise. Cross-infrastructure experiment can be constructed in that case to fulfil the requirements. In [12], a trans-oceanic integration of electrical equipment, controller and grid emulator is implemented. In [10] and [11], two frameworks (VILLAS and JanDER) connecting multiple hardware, real-time simulator and network simulator in the scale of Europe were presented. In a smaller scale, an integration of two experimental platforms PREDIS-PRISMES via hybrid cloud architecture is introduced in France [13], [14]. These novel approaches provide more versatility and flexibility to researchers to build the most suitable validation framework for their researches and also reinforce the international collaboration and exchange in the field of CPES research and development.

In this paper, we present our method of Supervisory, Control and Data Acquisition (SCADA) services integration in cross-infrastructure holistic tests of CPES. SCADA services provide a systematic approach and more complete information and control than a simple cross platform integration of single equipment. However, integration of different SCADA system is a complex process and requires rigorous consideration of different abstraction and technical layers (application, information model, communication protocol, etc.). The method is then illustrated with a case-study of combining assets and SCADA services of one platform to a remote HIL experiment in another platform, in the framework of a holistic CPES validation.

This work is supported by the Carnot Institute "Energies du Futur" framework under the PPInterop II project (www.energiesdufutur.eu). The participation of G2Elab and CEA-INES is also partially supported by the H2020 Erigrid project, Grant Agreement No. 654113.

The rest of the paper is organized as follows. Section II recalls the approach of cross-infrastructure holistic testing for smart grid and the interoperability requirements for such test. The architecture of hybrid cloud SCADA for integration of multiple RI in a holistic test is then presented in section III. In section IV, the experimental setup and results of a case study are shown to validate the proposed method. Section V concludes the paper and outlines possible future directions.

## II. CROSS-INFRASTRUCTURE HOLISTIC TESTING FOR SMART GRID AND INTEROPERABILITY REQUIREMENTS

CPES are multi-domain systems and due to their diverse and complex natures, several important remarks need to be considered in the implementation of its validation framework:

- The integrity and the interdependency of the subsystems in different domains need to be well reflected in the validation framework; either by specifying a co-simulation or by implementing a real integration (e.g. experiment with real communication system).

- A local solution obtained by a single domain test may not be (or compatible with) the global solution.

- The expertise or equipment necessary to fully characterize and implement a multi-domain validation framework may not be fully available in one RI and may require a combination of various RI (e.g. for a realistic assessment cyber-security in smart grid, it is required to have not only an extensive knowledge on ICT infrastructure (which is not always evident for the power system community), but also skills and installation of the electrical grid (which is in turn, not always available in an ICT RI)). Combining complementary expertise and platforms of RI in a holistic test is therefore a judicious solution that does not require heavy investments in time, infrastructure and human resources.

- Implementing cross-infrastructure holistic testing allows the RI to exchange specialized expertise and to get access to a shared resource pool, to manually or remotely borrow or share the infrastructure for research activities.

To successfully integrate different platforms together in a holistic test (e.g. remote HIL, co-simulation, or hybrid simulation), it is necessary and imperative to achieve interoperability among them. Moreover, interoperability promotes the possibility to visualize and to control in real time the available experimental tools, energetic systems and automation systems in the partner platforms. Achieving interoperability also facilitates the connection and integration of new platforms to the group and reduces the cost of installation of experimental modules and integration processes.

In [13], we proposed a five layers interoperability architecture among the RI (Figure 1). A holistic experiment integrates information and data from various sources with different formats. It is therefore necessary to achieve interoperability until information layer for an integration of SCADA service. Interoperability at function layer is required if one wants to integrate SCADA services from different RI into one single application. On the other hand, the harmonization of communication policies among RI is crucial in determining the application architecture and the model of service deliverance and integration, whether it is centralized or ad-hoc.

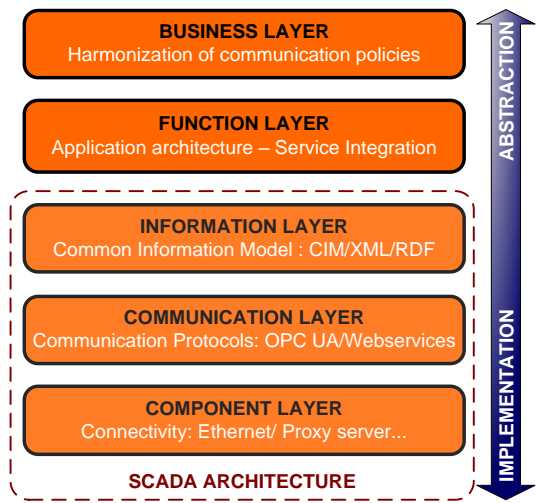

Figure 1: Interoperability architecture in cross-infrastructure holistic experiment.

In general, it is ideal (and best for performance) to achieve interoperability at all the concerned levels. However, due to practical situation, the RI involved in the holistic test may decide to keep their actual communication implementations. In that case, the deployment of a machine readable information model and experimental design is important to automate the orchestration and configuration of experiments. It is noteworthy that the "information model" layer here is in the sense of the equipment and the system configuration of the interested RI, and is not necessarily adapted to representing the semantic interaction among the test-elements, as well as the causality and the transition among test-stages. In the framework of a holistic test implementation, however, both need to be taken into account to fully assure the semantic coherence. As for the test-semantics in CPES, the testing description standard TTCN-3 [5] and the 62559 template from SGAM model [15] are notable options. However, no automated process has been registered up to now. On the other hand, several information models for interoperability in the electrical domain exist in the literature, in which CIM (IEC 61970 and IEC 61968) is considered to be a suitable option to configure the consortium data [16].

From an application point of view, holistic testing of a CPES involves using advanced experimental techniques such as real-time simulation and HIL, which requires the system configuration to satisfy rigorous real-time constraints, in terms of computational load, communication consistency and timing synchronicity. It is necessary to do a latency assessment in the implementation of the test, both in software/software and software/hardware interfaces, as well as in inter-RI interface. This assessment is especially important if the holistic test involves power equipment because the stability of the system is strongly influenced by the communication latency [17].

## III. SCADA-IN-THE-LOOP TESTING

The integration of SCADA systems provides a more systematic way to include various resources from different partners into a holistic experiment. Due to the distance among the participants RI, a hybrid cloud based approach was proposed in [13] to enable interoperability inter-platforms while ensuring the low latency, confidentiality and security for critical tasks. In this approach, the critical tasks are maintained on-site and are controlled by the local SCADA server, Programmable Logic Controller (PLC) or Remote Terminal Unit (RTU). Other applications such as Data Historian, Monitoring and Optimization can be done on a private cloud server with possibility to deliver these services to other partners via Platform-as-a-service (PaaS) or Software-as-a-service (SaaS) deliverance models. The interoperability among partners on this hybrid cloud server is actualized by applying CIM over lower level SCADA protocols (IEC 61850 or OPC UA). The harmonization of these different data models can be remedied and semi-automated with some tools developed in literature [18], [19]. We have also proposed an adaptive and automated method to deploy CIM over the consortium cloud, given the CIM/XML//RDF description of the participating RI in [20].

This architecture, besides the benefit of creating a mutual resource pool among the participant RI, can also be used as a base for integration and collaboration of distributed holistic test (Figure 2). As for the integration of HIL to co-simulation, it is necessary to create a software interface for the RTS and the hardware. For instance, we expect a common "hub" where the different simulators exchange the signals and the master algorithm executes synchronization. The hybrid cloud can be a suitable architecture for this software interface. In general, we can use a server based (cloud whiteboard) approach to integrate multiple resources to a holistic test-case (Figure 2). Interoperability at dynamic layer can be achieved.

Upon integration of SCADA service and implementation of a holistic test with this architecture, the system designer needs to take into account several considerations. While the partners can choose to adapt or to keep their information models, it is necessary to ensure a coherency of information model and communication protocols, at least at the cloud server. Besides, latency assessment and stability analysis of the interfaces need to be carefully done, especially in case of power equipment involvement.

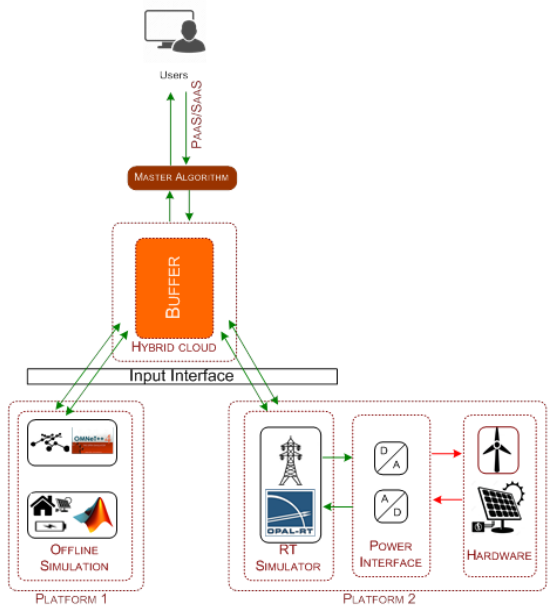

Figure 2: Integration RTS with other resources (SCADA, Simulator) via cloud server

## IV. IMPLEMENTATION AND RESULTS

In this section, to illustrate our approach, we apply this architecture to an implementation of a remote integration of hardware to RT simulator.

### A. Test-case description and cross-infrastructure set up

The proposed approach allows us to integrate assets from multiple platforms to a single holistic test-case and opens the door towards realistic validation of complex CPES. In order to demonstrate the applicability of the architecture, we consider in this section a test case of Distributed Multi-Agent System (MAS) control of IoT-based microgrid. The test-case is implemented by integrating Remote-HIL (SCADA server) in PRISMES platform (Le Bourget du Lac, France) with the RT Simulator (OPAL-RT) in PREDIS platform (Grenoble, France) - 70 km apart.

#### 1) Test-case description

The physical entities of AC microgrid comprises inverter-interfaced DGs (such as PV, wind turbines, energy storage systems), diesel generators, loads and etc. Each DG unit in the microgrid can be owned by the different stakeholders, and their controllers (including the control algorithm) may be geographically away from the microgrid entities. The MAS system in charge for secondary control of the microgrid is configured as proposed in [21]. The control architecture consists of three layers (Figure 3):

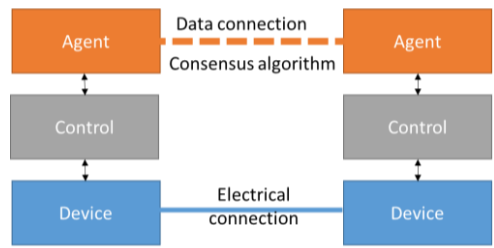

Figure 3: Proposed MAS based control architecture

Control agents are added on top of the local control layer. The role of the agent layer is to actualize optimization and to bring the microgrid state to desired conditions (e.g. regulation of voltage or frequency). The agents affect the grid via sending set-point to the local controllers. An agent only communicates with its neighbourhood and does not necessarily require knowing the state of the whole microgrid. The stability and consistency of the grid is ensured via a consensus algorithm among the agents. Due to this consensus process, the agents only send set-points to the primary controller once their outputs converge. In this particular case-study, we consider the regulation of frequency with the proposed MAS.

In order to validate this MAS structure of control in microgrid as well as demonstrate the functionality of our proposed architecture for integration of SCADA and HIL to co-simulation, the experiment in Figure 4 was carried out. The system under test in this experiment is a microgrid with 5 DRES (one PV Pack with inverter and 4 Energy Storage System ESS) and 2 loads (1 building and 1 other load).

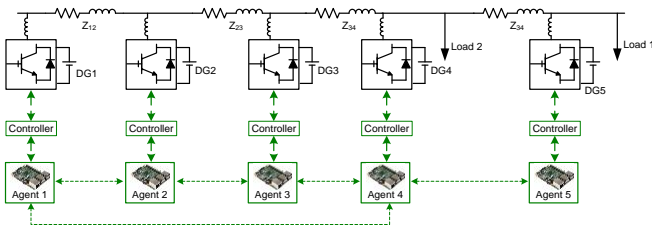

Figure 4: Microgrid under consideration

*2) Cross-infrastructure setup: Integration SCADA – RT simulator*

The smart building (Figure 5 – experimental building with advanced isolation, regulated temperature, automated water system, ventilation, lighting and human presence emulators) and the PV pack (20 Si-Monocrystalline panels and inverters) are physical hardware and are connected to the test via the OPC UA SCADA Panorama E2 of PRISMES platform in CEA.

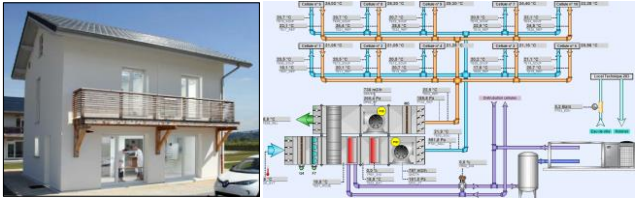

Figure 5: The smart building in consideration and its SCADA interface.

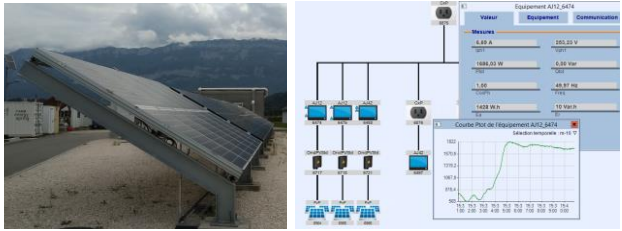

Figure 6: The considered PV Pack with inverter and its SCADA interface.

The agents are physical agents hosted on Raspberry PI cardboards and interconnected via a Netgear M4100-D12G hub, located at Grenoble INP. The communication inter-agent is then physical. The rest of system is simulated in OPAL-RT 5600. Cross-infrastructure communication is actualized via a Redis server. A Python interface is created to convert from OPC UA protocol to Redis protocol (Figure 7).

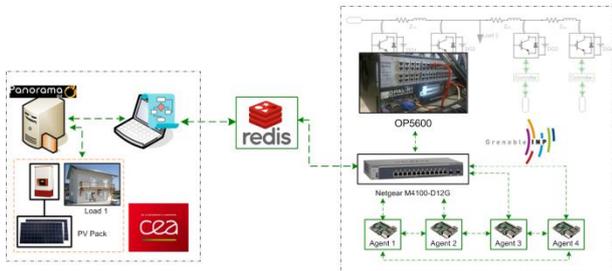

Figure 7: Implementation for the MAS control

A. *Experimental results*

The latency between the two platforms is measured to be consistently around 32 ms, with some occasional peaks of 85 ms (Figure 8). Due to this analysis, the SCADA system is set to emit measurements from the PV pack and the smart building every 1 second. In this configuration, we can make sure that the received signals are from the according emission step, and thus, well ordered.

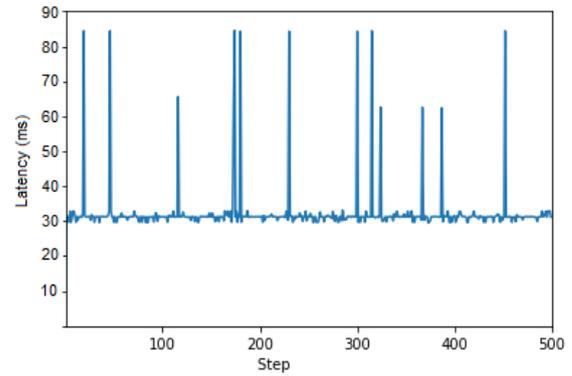

Figure 8: Latency between the OPC UA server and OPAL RT

The production curve of the considered PV pack in a typical summer day is given in Figure 9 (from 9 am to 4 pm on 30/07/2018). The simulation horizon of the rest of system (separately simulated in OPAL RT) must therefore intersect this period. In our test-case, we cropped the profiles in the model to start at 11 am and to end at 12h20 pm, i.e. the red frame in Figure 9. A similar procedure was actualized for the smart building consummation.

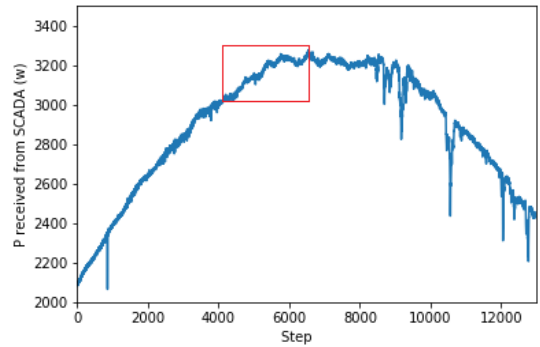

Figure 9: Production of the photovoltaic pack from 9 am to 4 pm.

The realistic PV production and the building consummation in CEA are synchronized with OPAL-RT at GINP (Figure 10). To investigate the functionality of the control algorithm, several disturbances were simulated: the PV production and the building consumption are connected to the microgrid at 11h00 am. Load 2 is connected to the grid after 60s and the building is disconnected from the microgrid at 120s. Against the three disturbances, the MAS successfully regulated the frequency of the system back to 50 Hz (Figure 11).

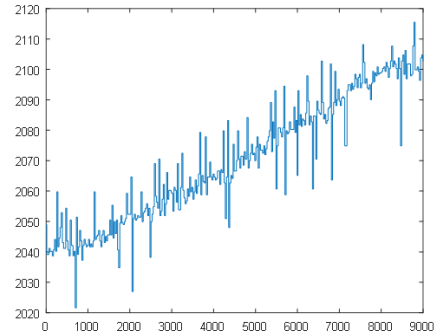

Figure 10: The PV pack production curve $P_{PV}$ (W) as injected to OPAL.

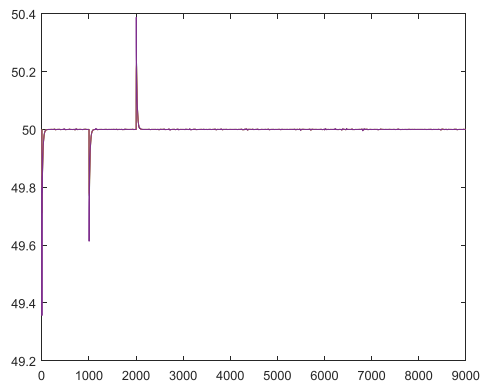

Figure 11: Successful regulation of frequency with the MAS network

This test-case has successfully demonstrated on one hand, the deployment of MAS control for IoT-based microgrid, on the other hand, the application of our proposed method of integration of SCADA system in a cross-infrastructure experiment of CPES, as interoperability architecture for combination of assets and expertise from various RI into a holistic test.

## V. CONCLUSION

CPES are complex multi-domain system of systems that requires sophisticated validation methodology and infrastructure. Combining assets and expertise of various RI in co-simulation or remote HIL is a promising approach towards such a validation framework. In this paper, we proposed the integration of SCADA services into cross-infrastructure experiments as a systematic way to integrate assets from different partners. The method is illustrated with a test-case of MAS control for IoT based microgrid. The implementation of the test-case involves the integration of SCADA system of CEA to the real-time simulation platform of Grenoble INP.